# Approximation of subsets of natural numbers by c.e. sets

Mohsen Mansouri[1], Farzad Didehvar[1,2]

**Abstract**

The approximation of natural numbers subsets has always been one of the fundamental issues in computability theory. Computable approximation, $\Delta_2$ − approximation. As well as introducing the generically computable sets have been some efforts for this purpose. In this paper, a type of approximation for natural numbers subsets by computably enumerable sets will be examined. For an infinite and non-c.e set, $W_i$ will be an $A. maximal$ (maximal inside A) if $W_i \subset A$ is infinite and $\forall j (W_i \subseteq W_j \subseteq A) \to \Delta(W_i, W_j) < \infty$, where $\Delta$ is the symmetric difference of the two sets. In this study, the natural numbers subsets will be examined from the maximal subset contents point of view, and we will categorize them on this basis. We will study c.regular sets that are non-c.e. and include a maximal set inside themselves, and c.irregular sets that are non-c.e. and non-immune sets which do not include maximal sets. Finally, we study the graph of relationship between c.e. subsets of c.irregular sets.

1. **Introduction**

In this paper, a type of approximation would be studied for natural numbers subsets by computably enumerable sets. Most methods in computability theory are trying to approximate the natural numbers subsets or examine how close a set is to being computable, so that after the computable sets, the most important sets are those that can be approximated by a computable sequence [1]. A computable approximation for approximating c.e. sets, $\Delta_2$-approximation, or trial and error method for approximating sets $A \leq_T 0'$ (when we have $A = \lim_s A_s$) as well as the $\Sigma_3$-approximation are some examples of these Methods.

It can be said that in [2], [3] and [4] observe this issue from another perspective. They study the relationship between classical asymptotic density and computability theory and examine the genericity concept in the computability theory and investigate the coverability of sets by computably enumerable sets which are generic.

**Definition 1.1-** If $A$ is a subset of $\omega = \{0,1,...\}$, and $n \geq 1$ then the density of $A$ up to $n$ is:

---

[1] Department of Mathematics and Computer Science, Amirkabir University of Technology,Tehran, Iran

[2] Corresponding author, Email address: didehvar@aut.ac.ir (Farzad Didehvar)

$$\rho_n(A) = \frac{|m \in A : m < n|}{n}$$

and the asymptotic density of A is $\rho(A) = \lim_{n \to \infty} \rho_n(A)$, if this limit exists.

**Definition 1.2-** $A$ is a generic set if $\rho(A) = 1$.

**Definition 1.3 [3]-** Let $\chi_A$ be the characteristic function of set A. It is a generically computable set if there exists a partial computable function $\varphi$ if whenever $\varphi(x) \downarrow$ then $\varphi(x) = \chi_A(x)$ and domain of $\varphi$ is a generic set. Simply put, $\varphi$ never gives an incorrect response about membership in A, though for some negligible inputs $\varphi$ gives no answer and doesn't converge.

Let $C(A) = \{2^n : n \in A\}$ For an arbitrary set $A \subset \mathbb{N}$. Then $C(A) \equiv_T A$ and $C(\bar{A})$ is a generically computable set. Also, with the definition of $R_k = \{m : 2^k | m, 2^{k-1} \nmid m\}$ and $\mathcal{R}(A) = \bigcup_{n \in A} R_n$ we have $\mathcal{R}(A) \equiv_T A$ and $\mathcal{R}(A)$ is generically computable if and only if A is a computable set. In other words, there is a non-generically computable set in each Turing degree [5].

[6] examines the question of how close a set can be to computable sets, while examining bi-immune and absolutely undecidable sets (or densely undecidable as suggested in [5]). A a set A extends the p.c. function $\varphi$ if $\varphi(x) \downarrow$ then $\varphi(x) = \chi_A(x)$, where $\chi_A$ is the characteristic function of A (note that it doesn't mention to the genericity of domain of $\varphi$). In this definition, the larger domain of $\varphi$ is, the set A is closer to being computable. Thus, in this regard, the bi-immune sets are the farthest sets to being computable because they do not extend any p.c. function with infinite domain.

For more information on the relationship between asymptotic density and computability theory, you can refer to [5].

Accordingly, the question that comes to mind is whether for a given infinite set $A$ there exists a c.e. set $W_i \subset A$ so that,

$$\forall j (W_j \subset A) \to W_j \subseteq W_i$$

In other words, does A have the largest c.e. subset?

Suppose that A is a non-computable set that $W_i \subset A$ and $W_i$ is the largest c.e. subset of A. Since $(A - W_i)$ is necessarily an infinite set, there is a $W_j = W_i \cup x$, such that $x \in (A - W_i)$, contradicts with $W_i$ for being maximum in A. Hence, finding the largest c.e. subset for A is impossible. For this reason, we define the Maximal set inside A as follows.

Definition ($A.maximal$): For a given set $A$, $W_i$ is $A.maximal$ (maximal inside A) if $W_i \subseteq A$ is infinite and

$$\forall j (W_i \subseteq W_j \subseteq A) \to \Delta(W_i, W_j) < \infty$$

Where $\Delta(W_i, W_j)$ is the symmetric difference of $W_i$ and $W_j$.

Note that if $W_i$ is $A.maximal$ for an arbitrary A, it is not necessarily a maximal set.

So, we ask our question like this: Let $A \subset \mathbb{N}$. Does $A$ have an $A.maximal$ set?

Immune sets are those that have no infinite c.e. subset. Therefore there is no maximal set inside them. Oppositely, each infinite c.e. set is a maximal inside itself. The next sections of the paper will examine two other types of natural numbers subsets in terms of the maximal existence inside them and are organized as follows.

In Section 2, first we introduce the c.regular sets that contain a maximal set and study the existence of these sets in c.e. degrees and below 0'. In Section 3, we will study the c.irregular sets and their Turing degrees. These type of sets have infinitely many c.e. subsets that none of which are a maximal inside the c.irregular set. In this section, we show that there are generically computable sets that there is no maximal inside them and hence are c.irregular. Finally, in Section 4, we study the relationship graph of the c.e. subsets of an c.irregular set.

## 2. c.regular sets

A Set $A$ is c.regular if $A$ is not c.e. and there is a maximal set inside it. More precisely,

**Definition 2-1:** A non-c.e. set $A$ is c.regular if there is a $W_i$ such that $W_i \subset A$ is infinite and

$$\forall j (W_i \subseteq W_j \subseteq A) \rightarrow \Delta(W_i, W_j) < \infty$$

So $W_i$ is A.maximal.

We will study the degree of these sets in the following. In Theorem 2-2 we show that for a given set $C \leq_T \mathbf{0'}$, we can computably in $C$ construct a c.regular set $A$, So we will use $\Delta_2^0\ permission$ method. In Theorem 2-3, for a given c.e. degree c, we construct set $A \in c$.

**Theorem 2-1:** For each set $C \leq_T \mathbf{0'}$, there is a c.regular set c.regular $A \leq_T C$

**Theorem 2-2.** In every c.e. degree there is a c.regular set.

## 3. c.irregular sets:

**Definition 3-1**: We consider a non-c.e. set $E$ as an c.irregular set, if for each $W_i \subset E$,

$$\exists j (W_i \subset W_j \subset E\ \&\ |W_j - W_i| = \infty)$$

The c.irregular sets are non- c.e. and non-immune sets which don't include maximal.

Probably, the most obvious example for the c.irregular sets are productive sets. For a productive set P, there is a computable and 1:1 function p, such that:

$$\forall x [W_x \subset P \rightarrow [p(x) \in P - W_x]]$$

Now if we consider $W_x \subset P$, for $W = \{p(x), ph(x), ph^2(x), ...\}$ we have $|W - W_x| = \infty$ and so on.

For each productive set P we have $P \geq_T 0'$. In theorem 3-1, we show that for each arbitrary set $C \leq_T 0'$, a c.irregular set can be constructed computably in C, and in theorem 3-2, we prove the existence of an

c.irregular set in each non-computable c.e. degree. At the end, we consider the relation of generically computable sets with c.regular and c.irregular sets.

**Theorem 3-2**: for $C \leq_T 0'$, there is an c.irregular set $E \leq_T$ C.

**Theorem 3-3**: In each non-computable c.e. degree c, there is an c.irregular set E.

**Definition 3-4**: a c.irregular set which includes a generically computable set is called generically c.irregular.

In the following corollaries we first show that although being generically computable means having an acceptable approximation, but this is not necessarily the best one, then we prove that including a maximal set doesn't mean to be generically computable.

**Corollary of theorem 3-2**: for each $C \leq_T 0'$ there is a generically c.irregular set like E, such that $E \leq_T$ C:

**Corollary of theorem 2-1**: for each $C \leq_T 0'$ set, there is a non-generically computable set like A, such that $A \leq_T C$.

## 4. The structure graph of c.irregular sets

In this section we study the structure and the relationship between c.e subsets of a c.irregular set. To do so, a directed graph $G_A = (V_A, E_A)$ will be considered for each c.irregular set A. Each vertex of $V_A$ includes a group of subsets of A, which have finite difference. In other words, if $v \in V_A$ and $B, C$ are c.e subsets of A, then

$$B, C \in v \leftrightarrow |\Delta(B, C)| < \infty$$

Also if $u, v \in V_A$ and $U \in u, V \in v$ then

$$(u, v) \in E_A \leftrightarrow |V - U| = \infty$$

$G_A$ is a strongly connected graph, in which there is at least one edge between every two arbitrary vertices u and v of $V_A$, and the indegrees and oudegrees of each vertex is infinite. An isomorphism of graphs of two c.irregular sets means they have same structure. we note that two graphs G and H are isomorph, if and only if there is a bijective function $f: V(G) \to V(H)$ between the vertices of these two graphs, such that $uv \in E(G)$ if and only if $f(u)f(v) \in E(H)$.

Question 4-1: do all c.irregular sets have an identical structure? In other words, are their graphs isomorphic?